# Protection of multilayer network systems from targeted group attacks

Olexandr Polishchuk, Dmytro Polishchuk

*Pidstryhach Institute for Applied Problems of Mechanics and Mathematics, National Academy of Sciences of Ukraine, Naukova str, 3"b", Lviv, 79060, Ukraine*
od_polishchuk@ukr.net

**Abstract**
The main types of simultaneous targeted group attacks on complex network systems and processes of intersystem interactions are discussed in the article. On the basis of structural model of multilayer network system (MLNS) and its aggregate-network, the most important components from a structural point of view, namely the cores of various types, whose damage will cause the greatest lesions of MLNS's structure, are highlighted. On the basis of flow model of multilayer system and its flow aggregate-network, the most important their components from a functional point of view, namely the flow cores of various types, whose damage will cause the greatest lesions of the process of inter-system interactions, are determined. Effective scenarios of successive and sim-ultaneous targeted group attacks on the structure and operation process of multilayer network systems have been developed using the structural and flow cores of aggregate-networks of MLNS. It is shown that the use of flow-based approach allows us to build much more effective scenarios of such attacks, as well as to more accurately evaluate the consequences of resulting lesions.

## 1. Introduction

The main types of negative internal and external influences on complex network systems (NS) and intersystem interaction processes were analyzed in the article [1]. Among these influences, targeted attacks and non-target lesions of complex systems were primarily highlighted, which can have local, group, or system-wide character and be aimed at damaging both the structure and operation process of network and multilayer network systems. The article also analyzed typical scenarios of sequential attacks on the structure and process of intersystem interactions, established their connection with the development of countermeasures against non-target system lesions, and proposed methods for evaluating the local and general losses caused by certain negative influence. No real-world large scale complex system is capable of simultaneously protecting or restoring all elements affected by negative influences [2, 3]. Currently, in the theory of complex networks (TCN), researchers' main focus is on constructing scenarios of sequential targeted attacks on the most structurally important elements of NS and MLNS [4, 5]. In monograph [6], based on structural and flow models of intersystem interactions, the main local and global structural and functional indicators of the importance of MLNS elements were identified, allowing for the detection of system elements that require primary protection. To reduce the problem's dimensionality, the concepts of structural and flow aggregate networks of MLNS were introduced, through which effective scenarios of sequential targeted attacks on the system's structure and operation process were constructed. It is evident that simultaneous group and system-wide attacks on NS and MLNS are significantly more dangerous, both in terms of their protection against lesion and their recovery afterward. For example, in Ukraine, the share of state banks in the country's banking system at the beginning of 2022 did not exceed 0.7%. At the same time, their share of assets in this system was 55.2%, and the share of individual deposits was 61.6% [7]. A successful attack on this small group of banks would lead to the greatest losses in the state's financial system. The massive DDoS attacks on January 14 and February 14-16, 2022, on more than 70 of Ukraine's most important state, security, financial, and social computer networks [8] can be considered an attempt at a system-wide strike on the information component of the state's governance system. This implies that to critically destabilize or shut down a real NS or MLNS, in many cases, it is enough to simultaneously damage the structure and/or operation process of a certain group of nodes. Indeed, sequential attacks on separate, even the most structurally important nodes of the network system, as proposed in currently developed targeted attack scenarios [9, 10], often allow us to redistribute their functions among other undamaged nodes. However, countering a simultaneous successful attack on a group of the most important elements of NS or MLNS, or the system as a whole, and, more importantly, overcoming the consequences of such attack or large-scale non-target lesion, is incomparably more difficult [11, 12]. The purpose of this article is to determine, based on structural and flow models of intersystem interactions, the importance indicators of MLNS components, develop effective scenarios of simultaneous group attacks on the structure and operation process of multilayer network systems, and evaluate the consequences of their damage for separate layer-systems and the implementation of intersystem interactions in general. Solving these problems will facilitate the correct decisions making not only regarding ensuring the active and passive protection of the system, but also organizing its recovery after lesions and the fastest possible return to normal operation.

## 2. The group lesions of complex network and multilayer network systems

A targeted attack or non-target lesion of even one of the most important elements of real-world system can lead to dangerous consequences (ranging from widespread dissatisfaction with the quality of information services to the declaration of war): the cyberattack on Kyivstar on December 12, 2024, difficulties with submitting electronic declarations in the spring of 2016 and 2017, the Chernobyl Nuclear Power Plant accident on May 26, 1986, the attack on Pearl Harbor on December 7, 1941, etc. Clearly, simultaneous group attacks or non-target system lesion can be much more difficult than point or sequential element-wise attacks, both in terms of system protection and overcoming the consequences. We categorize simultaneous group negative influence as one-time, repeated, and sequential. In the case of targeted attacks, this categorization is often determined by the attacker's ability to carry out subsequent mass attacks and the attacked system's capability to effectively defend against and counter them. Examples of one-time group negative influences include the terrorist attack by Al-Qaeda on the United States on September 11, 2001, which was carried out simultaneously on several civilian and military targets, and the Hamas missile attack on Israel on October 7, 2023, during which more than 2,500 rockets were launched. Repeated group attacks occur regularly over certain intervals on the same system targets. Examples of repeated attacks include the 18 missile strikes on Kyiv throughout May 2023, the continuous shelling of border and front-line settlements in Ukraine during the russian-ukrainian war, earthquakes and tsunamis in Japan and Chile, seasonal flu, waves of Covid-19, and more. Sequential group attacks differ from repeated ones by a change in the targets of damage: the series of missile attacks on Ukraine's oil depots in May-June 2022 and the airstrikes on Ukraine's power system transformer stations during 2022-2024, or the phased sanctions against russia's financial-economic system, defense-industrial complex, and so on. It is evident that each of the aforementioned types of attacks requires the development of specific scenarios for its most likely realization. The simplest scenario of one-time group attack is obviously realized by attempting to simultaneously strike a group of the most important MLNS elements, identified by a certain criteria. The scenario of repeated attack is realized by attempting to strike a previously selected and earlier attacked but not destroyed group of elements of the multilayer system. Scenario 1 of sequential group attack involves gradually executing the following steps:
1) create a list of groups of nodes (subsystems) of the MLNS in descending order of their structural and/or functional importance in the system;
2) remove the first group from the created list;
3) if the attack success criterion is met, end the scenario; otherwise, proceed to step 4;
4) since the system structure and operation process changes due to the removal of certain group of nodes (and their connections), create a new list of groups in descending order of recalculated structural and/or functional importance indicators in the MLNS, and return to step 2.

From the discussed above scenarios, it follows that in addition to determining the attack success criteria [6], the primary way to improve their effectiveness consist in selecting the structural and/or functional importance indicators of the group in the system, the lesion of which would cause the greatest harm. The most obvious way to make such selection is by forming a list of MLNS nodes in descending order of their structural or flow centrality of the chosen type and forming a group from the first nodes on this list, with the quantity of nodes determined by the intruder's ability to simultaneously attack them. The second approach is based on the principle of nesting hierarchy [13]. For example, before a military offensive, it is advisable to first destroy the command centers and key logistical objects of the enemy's army in the region adjacent to the front line where this offensive is planned, rather than those located far from it. If epidemic of a dangerous infectious disease begins in a certain area of the country, this area should be prioritized for isolation (quarantine). A similar situation arises in zones of radioactive or chemical contamination, areas of forest fires, or regions experiencing the proliferation of agricultural pests, etc. We will determine the importance indicators of MLNS groups of elements based on its structural and flow models and the concepts of aggregate-networks and cores of multilayer systems, which these models allow us to form.

## 3. A structural model of multilayer network system

The structural model of intersystem interactions is described by multilayer networks (MLNs) and displayed in the form [14]

$$G^M = \left( \bigcup_{m=1}^{M} G_m, \bigcup_{m,k=1,\, m \neq k}^{M} E_{mk} \right),$$

where $G_m = (V_m, E_m)$ determines the structure of $m^{\text{th}}$ network layer of MLN; $V_m$ and $E_m$ are the sets of nodes and edges of network $G_m$, respectively; $E_{mk}$ is the set of connections between the nodes of $V_m$ and $V_k$, $m \neq k$, $m,k = \overline{1,M}$, and $M$ is the quantity of MLN layers. The set

$$V^M = \bigcup_{m=1}^{M} V_m$$

will be called the total set of MLN nodes, $N^M$ – the quantity of elements of $V^M$. In this paper, we consider partially overlapped MLN [15], in which connections are possible only between nodes with the same numbers from the total set of nodes $V^M$ (Fig. 1a). This means that each node can be an element of several systems and perform one function in them, but in different ways. Nodes through which interlayer interactions are carried out will be called MLNS transition points, and the set

$$E^M = \bigcup_{m=1}^{M} E_m$$

is the total set of edges, $L^M$ – the quantity of elements of the set $E^M$.

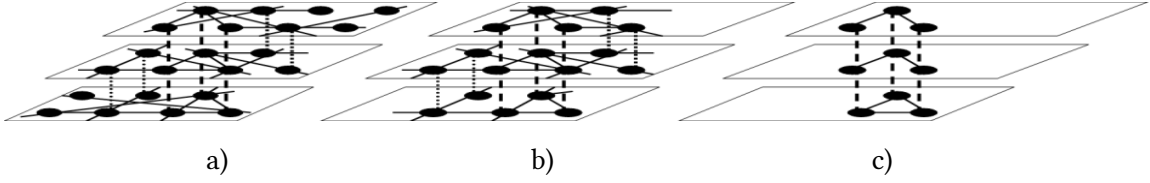

a)            b)            c)

**Figure 1:** Fragments of threelayer MLN (a) and its *2-* (b) and *3*-cores (c).

Multilayer network $G^M$ is fully described by an adjacency matrix

$$\mathbf{A}^M = \{\mathbf{A}^{km}\}_{m,k=1}^M, \qquad (1)$$

in which the blocks $\mathbf{A}^{mm}$ determine the structure of intralayer and blocks $\mathbf{A}^{km}$, $m \neq k$, – interlayer interactions. Values $a_{ij}^{km} = 1$ if the edge connected the nodes $n_i^k$ and $n_j^m$ exists, and $a_{ij}^{km} = 0$, $i, j = \overline{1, N^M}$, $m, k = \overline{1, M}$, if such edge don't exists. Blocks $\mathbf{A}^{km} = \{a_{ij}^{km}\}_{i,j=1}^{N^M}$, $m, k = \overline{1, M}$, of matrix $\mathbf{A}^M$ are determined for the total set of MLN nodes, i.e. the problem of coordination of node numbers is removed in case of their independent numbering for each layer.

In monograph [6], to simplify the analysis of MLNS structure and development of scenarios for sequential targeted attacks, the concept of its aggregate-network was introduced, which is fully described by an adjacency matrix

$$\mathbf{E} = \{\varepsilon_{ij}\}_{i,j=1}^{N^M}.$$

The off-diagonal elements $\varepsilon_{ij}$, $i \neq j$, of matrix $\mathbf{E}$ represent the structural aggregate weights of the edges $(n_i, n_j)$, i.e., the quantity of layers in which these edges are present. The diagonal elements $\varepsilon_{ii}$ correspond to the structural aggregate weights of nodes in the multilayer network, i.e., the quantity of layers to which these nodes belong, where $n_i$ and $n_j$, $i, j = \overline{1, N^M}$, are nodes from the total set of nodes $V^M$. The structure of this aggregate-network can be described as follows (Fig. 1a):

$$G_{ag}^M = (V^M, E^M). \qquad (2)$$

### 3.1. Structural cores of multilayer network system

To solve the problem of identifying the most structurally important components of intersystem interactions, we introduce the concept of *p*-core $\widetilde{G}^p = (\widetilde{V}^p, \widetilde{E}^p)$ of partially overlapped multilayer network, which is defined as its largest multilayer subnetwork, where the nodes belong to at least *p*, $2 \leq p \leq M$, layers (Fig. 1b, c). The structure of *p*-core is described by an adjacency matrix $\widetilde{\mathbf{A}}_p^M$, which is derived from the adjacency matrix $\mathbf{A}^M$ by removing those rows and columns where the aggregate weight of nodes are less than *p*. If the maximum value *p*, at which the partially overlapped multilayer network $\widetilde{G}^p$ does not degenerate into an empty set, is equal to *M*, we will call such MLN coreness; otherwise, it is coreless. Clearly, the core $\widetilde{G}^M$ of coreness MLN has a multiplex structure (Fig. 1c) [16].

Elements of matrix $\mathbf{E}$ define the integral structural characteristics of the nodes and edges of multilayer network (Fig. 2a). The projections of *p*-cores, $2 \leq p \leq M$, onto the aggregate-network $G_{ag}^M$ will be called $p_{ag}$-cores of this aggregate-network (Fig. 2b, c).

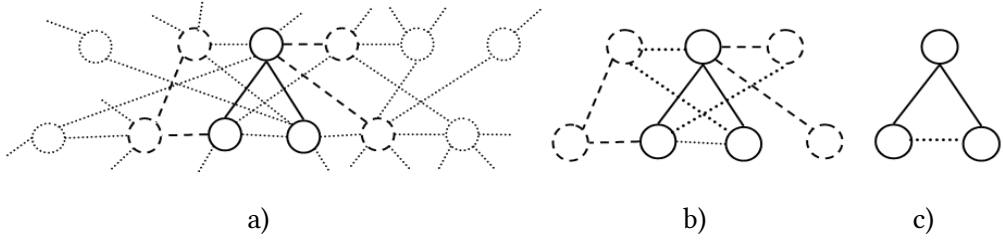

a)                          b)              c)

**Figure 2:** Fragments of structural aggregate-network of threelayer coreness MLN (a) and its $2_{ag}$- (b) and $3_{ag}$-cores (c) (___ – element belong to three layers, - - - – element belong to two layers, ..... – element belong to one layer).

The structure of $p_{ag}$-cores is described by adjacency matrices $\mathbf{E}_p$, which are derived from matrix $\mathbf{E}$ by removing rows and columns whose aggregate weights of diagonal elements are smaller than the value of $p_{ag}$. Identifying the $p$-cores and $p_{ag}$-cores, $2 \leq p, p_{ag} \leq M$, is one of the ways to recognize the most structurally important groups of nodes for organization of intersystem interactions in a partially overlapped multilayer network, which can become the primary targets for attacks on such interactions.

To highlight the most important components of complex network, the concept of its k-core is introduced, defined as the largest subnetwork of the source network whose structural degree of nodes is no less than k > 1 [17]. The analogous concept in multilayer network is so-called **k**-core [18], represented as a vector $\mathbf{k} = \{k_1, k_2, ..., k_M\}$, as combination of $k_m$-cores of separate MLN layers. In this case, the values $k_m$, $m = \overline{1, M}$, may vary across different layers. Generally, the **k**-core defines components that are structurally important for the MLN layers, but not for the organization of interlayer interactions within it, and is used to analyze so-called multidimensional (multiflow) MLNs [19]. For monoflow multilayer networks, which are considered in this article, we can introduce the concept of k-core as the largest multilayer subnetwork of the source MLN, whose generalized structural degree of nodes (the sum of quantities of input and output edges of nodes in network layers and its interlayer links at transition point [6]) is no less than k. Unlike the $p$-cores of MLN, their k-cores highlight the most important groups of nodes for both intersystem and intrasystem interactions in a partially overlapped multilayer network, which may become the primary targets for attack. We will call the projection of MLN k-core onto the aggregate-network $G_{ag}^M$ its $k^{ag}$-core. The structures of k- and $k^{ag}$-cores are described by the adjacency matrices $\mathbf{A}_k^M$ and $\mathbf{E}^k$, which are obviously derived from the matrices $\mathbf{A}^M$ and $\mathbf{E}$, respectively.

### 3.2. Targeted group attacks on MLNS structure

It should be noted that when identifying $p$-, $p_{ag}$-, k-, or $k^{ag}$-cores in the structures of real-world MLNs, a large quantity of disconnected groups of nodes included in these cores may appear. Then raises the issue of determining the importance indicators of these groups in MLN or its aggregate-network to form appropriate lists for scenarios of targeted simultaneous group attacks. To determine such importance indicators, one can use:

- the specific weight of the quantity of nodes in the group in the total set of nodes $V^M$;
- the specific weight of the quantity of edges between the nodes of the group in the total set of edges $E^M$;
- the specific weight of transition points of the group in the total set of transition points of multilayer network.

The importance of group in the MLN can also be determined by its generalized structural degree. In fact, the generalized structural degree of group determines the quantity of MLN nodes that can be consequentially injured as a result of simultaneous attack on this group. The sum of directly damaged and consequentially injured nodes due to such attack on the multilayer network can be considered the most suitable structural indicator of the group's importance. Based on these considerations and using, for example, the concept of k$^{ag}$-core, we can formulate *Scenario 2* of sequential targeted simultaneous group attack on the MLN structure:
1) set the value $q$ = max {k$^{ag}$};
2) create a list of groups of nodes that are part of $q$-core in MLN aggregate-network;
3) sort the compiled list of groups in descending order according to the selected importance indicator in aggregate-network, for example, the generalized structural degree of the group;
4) remove the first group from sorted list;
5) if the attack success criterion is met, terminate the execution of scenario; otherwise, proceed to point 6;
6) if the list of groups with current value of $q$ is not exhausted, return to point 3; otherwise, proceed to point 7;
7) set $q=q-1$; if $q$ is less than the minimum k$^{ag}$ value, terminate the execution of scenario; otherwise, return to point 2.

If during the execution of Scenario 2, a certain group of nodes contains too many elements for the attacker to target simultaneously, that group should be divided into minimum quantity of connected subgroups available for such attacks. Additionally, the scenario may end when the attacker's resources for continuing the attack are exhausted. It should be noted that as the value of $q$ sequentially decreases in the above scenario, the group attack gradually evolves into a system-wide attack.

## 4. A flow model of multilayer network system

A method for decomposing multidimensional MLNS into monoflow multilayer systems was proposed, and a flow model of these systems was considered in the article [1], which allows us to calculate the main local and global functional characteristics of elements of such formations and construct scenarios of successive group attacks on the process of intersystem interactions. By the flow on an edge, we mean a certain positive real-valued function associated with this edge (e.g., the number of passengers or tons of cargo transported between two neighboring stations per day, the quantity of cars that drove between two adjacent intersections of a city street per hour, the volume of natural gas that passed between two distribution stations during a month, the volume in kilobytes of a letter sent from one email user to another, etc.). Let us reflect the set of flows that pass through all edges of multilayer system in the form of flow adjacency matrix $\mathbf{V}^M(t)$, the elements of which are determined by the volumes of flows that passed through the edges of MLN (1) for the period $[t-T, t]$ up to the current moment of time $t \geq T$:

$$\mathbf{V}^M(t) = \{V_{ij}^{km}(t)\}_{i,j=1}^{N}{}_{k,m=1}^{M}, \quad V_{ij}^{km}(t) = \tilde{V}_{ij}^{km}(t) \Big/ \max_{s,g=1,M} \max_{l,p=1,N^M} \{\tilde{V}_{lp}^{sg}(t)\}, \quad (3)$$

where $\tilde{V}_{ij}^{km}(t)$ is the volume of flows that passed through the edge ($n_i^k, n_j^m$) of multilayer network for the time period $[t-T, t]$, $i, j = \overline{1, N^M}$, $k, m = \overline{1, M}$, $t \geq T > 0$, [6]. It is obvious that structure of matrix $\mathbf{V}^M(t)$ completely coincides with the structure of matrix $\mathbf{A}^M$. The elements of MLNS flow adjacency matrix are determined on the basis of empirical data about movement of flows through MLNS edges. Currently, with the help of modern means of information extraction, such data can be easily obtained for many natural and the vast majority of man-made systems [20]. The matrix $\mathbf{V}^M(t)$ similarly to $\mathbf{A}^M$ also has a block structure, in which the diagonal blocks $\mathbf{V}^{mm}(t)$ describe the volumes of intralayer flows in the $m^{th}$ layer, and the off-diagonal blocks $\mathbf{V}^{km}(t)$, $m \neq k$, describe the volumes of flows between the $m^{th}$ and $k^{th}$ layers of MLNS, $m, k = \overline{1, M}$, $t \geq T > 0$.

To identify the functionally most important components of monoflow multilayer system, introduce the concept of its flow $\lambda$-core. The adjacency matrix $\mathbf{V}_\lambda^M(t)$ of this core is determined from the model (3) by the relation:

$$\mathbf{V}_\lambda^M(t) = \{V_{\lambda,ij}^{km}(t)\}_{i,j=1}^{N^M}\,_{k,m=1}^{M}, \quad V_{\lambda,ij}^{km}(t) = \begin{cases} V_{ij}^{km}(t), \text{if } V_{ij}^{km}(t) \geq \lambda, \\ 0, \text{if } V_{ij}^{km}(t) < \lambda, \end{cases}$$

$$\lambda \in [0,1],\ t \geq T > 0,\ i,j = \overline{1,N^M},\ k,m = \overline{1,M}.$$

It is evident that the larger the value $\lambda$, the more functionally significant component of the multilayer system represented by its flow $\lambda$-core. This core may become one of the primary targets for simultaneous group attacks.

The concept of flow aggregate-network of monoflow partially overlapped MLNS was introduced in monograph [6]. Since we are considering the case when interlayer connections are possible only between nodes with the same numbers in total set of MLNS nodes, the structure of such aggregate-network can also be described in the form (2). Then the adjacency matrix

$$\mathbf{F}(t) = \{f_{ij}(t)\}_{i,j=1}^{N^M},$$

the elements of which are calculated according to the formulas:

$$f_{ij}(t) = \sum_{m=1}^{M} V_{ij}^{mm}(t)/M,\ i \neq j,$$

$$f_{ii}(t) = \sum_{m,k=1,\,m\neq k}^{M} V_{ii}^{mk}(t)/(M-1)^2,\ i,j = \overline{1,N^M},$$

completely defines a dynamic (in the sense of dependence on time) weighted network, which will be called the flow aggregate-network of this MLNS. The elements of matrix $\mathbf{F}(t)$ determine the integral flow characteristics of the edges and transition points of multilayer system, namely, the off-diagonal elements of this matrix are equal to the total volumes of flows passing through the edge $(n_i, n_j)$, and the diagonal elements are equal to the total volumes of flows passing through the transition point $n_i$ of MLNS during the time period $[t-T, t]$, $t \geq T > 0$, where $(n_i, n_j)$ are the edges from the total set of edges $E^M$, and $n_i$, $n_j$, $i,j = \overline{1,N^M}$, are the nodes from the total set of nodes $V^M$.

To identify the functionally most important components of the MLNS flow aggregate-network we introduce the concept of its flow $\lambda_{ag}$-core, whose adjacency matrix is determined by the following relation:

$$\mathbf{F}_{\lambda_{ag}}(t) = \{f_{ij}^{\lambda_{ag}}(t)\}_{i,j=1}^{N^M},\quad f_{ij}^{\lambda_{ag}}(t) = \begin{cases} f_{ij}(t), \text{if } f_{ij}(t) \geq \lambda_{ag}, \\ 0, \text{if } f_{ij}(t) < \lambda_{ag}, \end{cases} \lambda_{ag} \in [0,1],\ t \geq T > 0.$$

It is evident that the larger the value $\lambda_{ag}$, the more functionally significant component of the MLNS flow aggregate-network represented by its $\lambda_{ag}$-core. It is also advisable to select this core as one of the primary targets for simultaneous group attacks. It should be noted that the structures of projection onto the aggregate-network $\lambda$-core and the $\lambda_{ag}$-core, for equal values of $\lambda$ and $\lambda_{ag}$, generally differ, and the $\lambda_{ag}$-core determines the integral importance indicators of MLNS components.

## 5. Importance indicators for flow cores of multilayer network systems

The global flow characteristics of MLNS nodes, such as their input and output influence and betweenness parameters were introduced in monograph [6]. These parameters allow us to determine the importance of separate nodes in operation of multilayer system as generators, final receivers, and transitors of flows, develop the effective scenarios of targeted sequential element-wise attacks on the process of intra- and intersystem interactions. However, to form the effective scenarios of simultaneous group attacks, it is advisable to calculate the functional importance indicators of separate MLNS subsystems. To simplify the presentation, we define such indicators for $\lambda_{ag}$-core of MLNS aggregate-network.

### 5.1. Influence parameters of flow cores

Let us set the value $\lambda_{ag} \in [0,1]$ and denote by $H_{\lambda_{ag}} = \{n_i^{\lambda_{ag}}\}_{i=1}^{N_{\lambda_{ag}}}$ the set of nodes in $\lambda_{ag}$-core of multilayer system aggregate-network. Denote by $G_{\lambda_{ag}}^{out}$ the set of all nodes-generators of flows that belong to $H_{\lambda_{ag}}$ and $R_{\lambda_{ag}}^{out}$ is the set of indices of nodes that are the final receivers of flows generated by the nodes belonging to $G_{\lambda_{ag}}^{out}$. Divide the set $R_{\lambda_{ag}}^{out}$ into two subsets:

$$R_{\lambda_{ag}}^{out} = R_{\lambda_{ag},int}^{out} \cup R_{\lambda_{ag},ext}^{out},$$

where $R_{\lambda_{ag},int}^{out}$ is the subset of indices of nodes from $R_{\lambda_{ag}}^{out}$ that belong to $H_{\lambda_{ag}}$, and $R_{\lambda_{ag},ext}^{out}$ is the subset of indices of nodes from $R_{\lambda_{ag}}^{out}$ that belong to the complement of $H_{\lambda_{ag}}$ in the source aggregate-network. The set $R_{\lambda_{ag},ext}^{out}$ will be called the domain of output influence of $\lambda_{ag}$-core onto the MLNS flow aggregate-network, and the quantity of elements $p_{\lambda_{ag},ext}^{out}$ in this set – the power of this influence.

The external and internal output strength of influence of the nodes-generators of flows belonging to the set $G_{\lambda_{ag}}^{out}$ on the subnetworks $R_{\lambda_{ag},ext}^{out}$ and $R_{\lambda_{ag},int}^{out}$ will be calculated using the parameters:

$$\xi_{\lambda_{ag},ext}^{out}(t) = \sum_{i \in R_{\lambda_{ag},ext}^{out}} \xi_i^{out}(t) / p_{\lambda_{ag},ext}^{out},$$

$$\xi_{\lambda_{ag},int}^{out}(t) = \sum_{i \in R_{\lambda_{ag},int}^{out}} \xi_i^{out}(t) / p_{\lambda_{ag},int}^{out}, \qquad (4)$$

respectively. In formulas (4) the value $\xi_i^{out}(t)$ determines the total volume of flows generated in the node $n_i \in G_{\lambda_{ag}}^{out}$, that is, the influence strength of this node on the flow aggregate-network of multilayer system [6], and the value $p_{\lambda_{ag},int}^{out}$ is equal to the quantity of elements in the subset $R_{\lambda_{ag},int}^{out}$. The parameters $\xi_{\lambda_{ag},ext}^{out}$, $R_{\lambda_{ag},ext}^{out}$, and $p_{\lambda_{ag},ext}^{out}$ will be called the output parameters of influence of $\lambda_{ag}$-core onto the flow aggregate-network. Similarly are determinated the parameters $\xi_{\lambda_{ag},ext}^{in}$, $R_{\lambda_{ag},ext}^{in}$, and $p_{\lambda_{ag},ext}^{in}$ of input influence of the MLNS aggregate-network onto its $\lambda_{ag}$-core, i.e., the set of nodes-generators of flows outside this core in the MLNS aggregate-network on the nodes – final receivers of flows within $\lambda_{ag}$-core. The lesion of node-generator of flows means that the nodes – final receivers must find new sources of supply, while the damage of node – final receivers means that producers must find new markets, leading to at least temporary difficulties in their operations. The input and output influence parameters of $\lambda_{ag}$-core make it possible to quantify the losses resulting from a successful simultaneous attack on it and how far and to what extent such attack will spread across the elements of intra- and intersystem interactions.

## 5.2. Betweenness parameters of flow cores

No less important for the analysis of participation the $\lambda_{ag}$-core in operation process of MLNS flow aggregate-network are the betweenness parameters of this core which determine as follows. Denote by $P_{\lambda_{ag}}^{K_{\lambda_{ag}}} = \{p_{\lambda_{ag}}^k\}_{k=1}^{K_{\lambda_{ag}}}$ the set of paths that connect the generator nodes and final receiver nodes of aggregate-network flows, which lie outside the $\lambda_{ag}$-core, but pass through the elements of the set $H_{\lambda_{ag}}$. Let $v_{\lambda_{ag}}^k(t)$ be the volume of flows that passed through path $p_{\lambda_{ag}}^k$ from the generator node to the final receiver node, and therefore through $\lambda_{ag}$-core, during the period $[t-T, t]$. Then the value

$$V_{\lambda_{ag}}^{K_{\lambda_{ag}}}(t) = \sum_{k=1}^{K_{\lambda_{ag}}} v_{\lambda_{ag}}^k(t)$$

determines the total volume of flows that passed through the set of paths $P_{\lambda_{ag}}^{K_{\lambda_{ag}}}$, and therefore through $\lambda_{ag}$-core, during the same period of time. The value

$$\Phi_{\lambda_{ag}} = V_{\lambda_{ag}}^{K_{\lambda_{ag}}}(t) / s(\mathbf{F}(t)), \qquad (5)$$

which determines the specific weight of flows transiting through the $\lambda_{ag}$-core for period $[t-T, t], t \geq T$, will be called the measure of betweenness of this core in operation process of MLNS aggregate-network. In formula (5), the value $s(\mathbf{F}(t))$ is equal to the sum of elements of the flow adjacency matrix $\mathbf{F}(t)$. The set $M_{\lambda_{ag}}$ of all aggregate-network nodes that lie on paths from the set $P_{\lambda_{ag}}^{K_{\lambda_{ag}}}$ outside the $\lambda_{ag}$-core will be called the betweenness domain, and the quantity $\eta_{\lambda_{ag}}$ of these nodes will be called the betweenness power of $\lambda_{ag}$-core in the operation process of MLNS aggregate-network. The parameters of measure, domain and power of betweenness of $\lambda_{ag}$-core are global characteristics of its importance in the operation process of multilayer system aggregate-network. They determine how the blocking of this core will affect on the work of betweenness domain, the size of this domain and, as a result, the entire system.

## 5.3. Comprehensive scenario of targeted group attack

As in the case of nodes of MLNS aggregate-network [6], the values of influence and betweenness parameters of $\lambda_{ag}$-core can be generalized, taking into account that it can simultaneously be a generator, final receiver and transitor of flows. Namely, the generalized parameter $\Xi_{\lambda_{ag}}(t)$ of the force of interaction of $\lambda_{ag}$-core with MLNS aggregate-network in general, which is calculated by the formula:

$$\Xi_{\lambda_{ag}}(t) = (\xi_{\lambda_{ag},ext}^{out}(t) + \xi_{\lambda_{ag},ext}^{in}(t) + \Phi_{\lambda_{ag}}(t))/3, \quad t \geq T,$$

defines the overall role of $\lambda_{ag}$-core in aggregate-network of multilayer system as generator, final receiver and transitor of flows; the domain $\Omega_{\lambda_{ag}}(t)$ of interaction of $\lambda_{ag}$-core with MLNS aggregate-network is determined by the ratio:

$$\Omega_{\lambda_{ag}}(t) = R_{\lambda_{ag},ext}^{in}(t) \bigcup R_{\lambda_{ag},ext}^{out}(t) \bigcup M_{\lambda_{ag}}(t),$$

and the power $\eta_{\lambda_{ag}}$ of interaction of $\lambda_{ag}$-core with MLNS aggregate-network is equal to the ratio of quantity of elements of domain $\Omega_{\lambda_{ag}}(t)$, $t \geq T$, to the value $N^M$. It is clear that parameters of interaction of flow $\lambda_{ag}$-core with MLNS determine the level of their dependence on each other and make it possible to quantitatively define how damage to this core will affect the process of intersystem interactions in general, how many and exactly which elements of the aggregate-network of multilayer system will be affected and to what extent [21-23]. That is, in the case of lesion of $\lambda_{ag}$-core, the domain $\Omega_{\lambda_{ag}}(t)$ determines the totality of all consequentially injured MLNS elements, and parameter $\eta_{\lambda_{ag}}$ – their number. By means of concept $\lambda_{ag}$-core and generalized parameter $\Xi_{\lambda_{ag}}(t)$ of the strength of their interaction with MLNS flow aggregate-network as importance indicator of group of nodes, we can form *Scenario 3* of successive targeted simultaneous group attack on the process of intersystem interctions:
1) set the value $\lambda_{ag} = 1$;
2) compile a list of unconnected groups of nodes that are part of $\lambda_{ag}$-core in the MLNS flow aggregate-network;
3) sort the compiled list of groups in decreasing order based on the strength of their interaction with MLNS aggregate-network;
4) remove the first group from the sorted list;
5) if the attack success criterion is met, terminate the execution of scenario; otherwise, proceed to step 6;
6) if the list of groups with the current value $\lambda_{ag}$ is not exhausted, return to step 3; otherwise, proceed to step 7;
7) set $\lambda_{ag} = \lambda_{ag} - \delta$, where $\delta \ll 1$, $\delta \in [0,1]$ is a predefined value, for example, $\delta = 0,1$; if $\lambda_{ag}$ is less than its minimum value for the flow aggregate-network, terminate the execution of scenario; otherwise, return to step 2.

If during the execution of *Scenario 3*, a certain group of nodes contains too many elements that the attacker is unable to target simultaneously, such group is divided into the minimum number of connected subgroups accessible for such attacks. Additionally, the scenario may terminate when the intruder runs out of resources to continue the attack. It should be noted that as the value $\lambda_{ag}$ decreases in the above scenario, the group attack increasingly transforms into a system-wide attack [24-26].

Depending on the goal of attack, the targets may include generators, final receivers, flow transitors, or only transition points of $\lambda_{ag}$-core of the MLNS flow aggregate-network. For each of these types of nodes, specific targeted attack scenarios can be constructed, using the influence or betweenness parameters defined above in formulas (4) or (5), respectively, as indicators of group importance. One of the drawbacks of targeted attack scenarios based on local structural or functional importance indicators of MLNS nodes is that only the elements directly adjacent to the damaged nodes can reasonably be considered consequentially injured. Before conducting an attack on generators, final receivers, transitors, or transition points of MLNS, it is possible to identify the domains of input and output influence and betweenness, which help determine the elements that may be consequentially injured as a result of the attack, as well as to calculate the potential level of their losses. A quantitative measure of these losses relative to the damage inflicted on attacked system allows us to determine the feasibility of conducting the attack, for example, the imposition of specific sanctions against an aggressor country.

Similarly, as for the $\lambda_{ag}$-core, functional importance indicators and corresponding scenarios can be formed for arbitrary, e.g., hierarchically nested MLNS subsystems [13], connected groups of aggregate-network elements, or the $\lambda$-core of multilayer system as a whole.

# 6. Comparison of structural and flow-based scenarios of targeted group attacks

Let us consider the railway transport system (RTS) of the western region of Ukraine as an example of component of multilayer general transportation system of the country. This MLNS includes railway, automotive, water, and aviation layers. The structural model of RTS is built based on the railway connection map of the region (in this case, it includes 354 nodes). To develop a flow model, we use data about freight transportation volumes carried by rail during 2021 (in the next years, access to such data has been significantly restricted for understandable reasons). For better comprehension, fig. 3a shows the structural model of this network system without transit nodes of degree 2. This model includes 29 nodes and 62 edges. Fig. 3b illustrates a weighted network schematically reflecting the flow volumes that passed through the RTS edges during specified period (the line thickness is proportional to these flow volumes). Fig. 3c presents the structural 4-core of this network system, which includes 12 nodes and 35 edges, while fig. 3d shows its flow *0.8*-core, comprising 4 nodes and 12 edges. From the presented figures, one can observe a major drawback of k-cores: they may exclude functionally important system components from its structure (e.g., the path A–B).

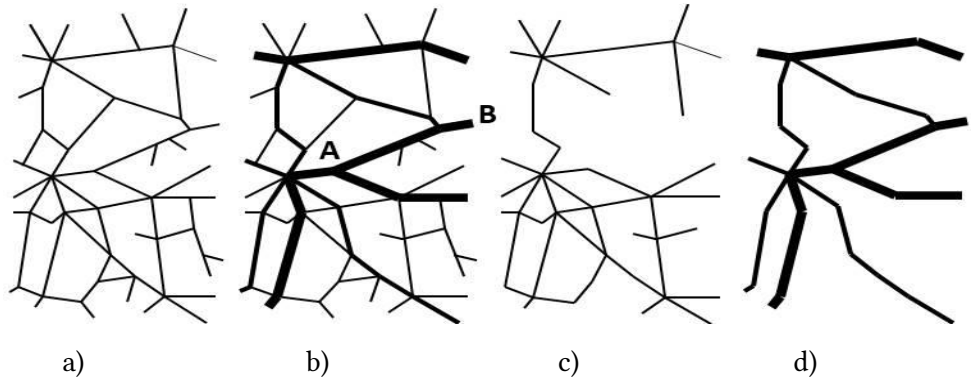

a)  b)  c)  d)

**Figure 3:** Examples of structure and operation process of railway transport system of the Western region of Ukraine (freight transportation).

The provided examples indicate that the quantity of targets in a group attack scenario based on the concept of flow core is three times smaller than in a scenario using k-core concept. Analyzing the influence and betweenness parameters of directly damaged nodes (attack targets) for presented network system indicates that all RTS elements are affected in this case. Thus, the flow-based approach enables the development of significantly more efficient group attack scenarios in terms of the number of attack targets, causing no less damage than the structural approach [27-29].

Nodes of transportation network that facilitate movement of the largest volumes of flows within the system require priority protection from targeted attacks. At the same time, during the spread of epidemics caused by dangerous infectious diseases, such nodes need to be promptly isolated to block passenger traffic. Thus, blocking separate MS components can serve both as an attack goal and as a method of system protection. Consequently, the problem of system vulnerability can be conditionally divided into two tasks. The first of them, discussed in the previous example, involves identifying the elements that need to be prioritized for protection to prevent system destabilization or operational failure. The second task focuses on determination the elements whose blocking would minimize the losses expected from the spread of lesion. We will demonstrate, using the example of railway passenger transportation system, that scenarios designed to protect the NS from targeted attacks can be effectively applied to counteract the spread of nontarget lesions. It is evident that the structural model of passenger transportation system, excluding nodes of degree 2 (fig. 4a), and its 4-core, coincide with the structural model of the freight transportation system. To develop a flow model for passenger movement, we use data on passenger traffic volumes handled by the railway in 2019 (prior to beginning the Covid-19 pandemic). Fig. 4b schematically illustrates this model (as before, the line thickness is proportional to flow volumes). Fig. 4d shows the flow *0.8*-core of passenger transportation system, containing 3 nodes and 8 edges. This indicates that halting passenger traffic requires blocking 4 times fewer elements compared to using the structural 4-core of corresponding network system.

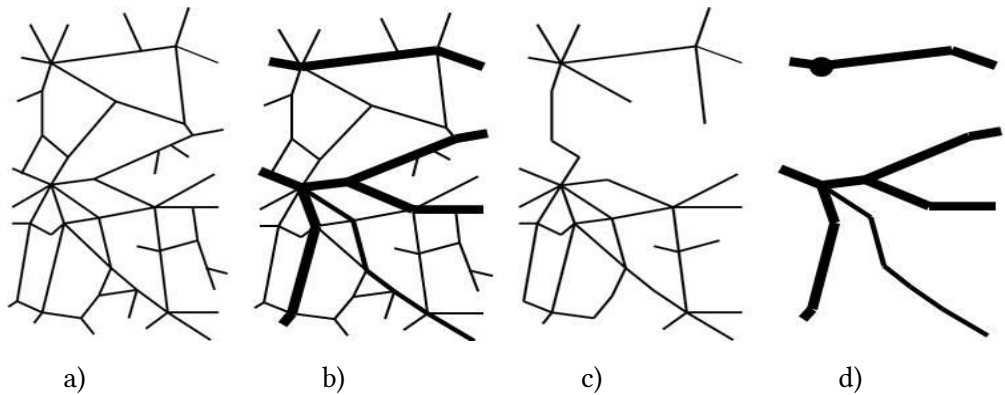

a)   b)   c)   d)

**Figue 4:** Examples of structure and operation process of railway transport system of the Western region of Ukraine (passenger transportation).

## 7. Conclusions

The main types of simultaneous targeted group attacks on complex network systems and intersystem interaction processes are considered in the article. Based on the structural model of multilayer network system and its aggregate-network, the most important structural components, namely, cores of various types, were identified, the disruption of which would cause the greatest damage to the MLNS structure. Based on the flow model of multilayer system and its flow aggregate-network, the most functionally important components were determined, specifically flow cores of different types, the disruption of which would cause the greatest lesions of intersystem interaction process. Using the structural and flow cores of MLNS aggregate-networks, effective scenarios of targeted simultaneous group attacks on the structure and operation process of multilayer network systems were developed. It is shown that application of the flow-based approach allows us to create the significantly more effective attack scenarios and a more accurate evaluate the caused by attack damage consequences. The next steps of our research are development the methods of system-wide attacks on complex network systems and intersystem interaction processes, analysis the problem of the scale of consequences from targeted attacks and non-target lesions, and creation the methods for optimizing counteraction scenarios against various negative influences on multilayer network systems.